\documentclass[aps,prl,twocolumn,superscriptaddress,floatfix,nofootinbib,showpacs,longbibliography]{revtex4-2}

\usepackage[utf8]{inputenc}  
\usepackage[T1]{fontenc}
\usepackage[british]{babel}
\usepackage[sc,osf]{mathpazo}
\usepackage[table]{xcolor}
\usepackage[scaled=0.86]{berasans}
\usepackage[colorlinks=true, citecolor=blue, urlcolor=blue]{hyperref}
\usepackage{graphicx}
\usepackage[babel]{microtype}
\usepackage{amsmath,amssymb,amsthm,bm,amsfonts,mathrsfs,bbm,physics,svg}

\usepackage{xspace}
\usepackage{pgf,tikz}
\usepackage{multirow}
\usepackage{array}
\usepackage{bigstrut}
\usepackage{braket}
\usepackage{mathtools}
\usepackage{float}
\usepackage[caption=false]{subfig}
\usepackage{color}
\usepackage{natbib}
\usepackage[normalem]{ulem}
\usepackage{dsfont}

\newtheorem{theorem}{Theorem}
\newtheorem{lemma}{Lemma}

\newtheorem{definition}{Definition}

\begin{document}

\title{Exclusive Control of Quantum Memory Erasure}

\author{Mir Alimuddin}  \thanks{These authors contributed equally and share first authorship. \\ aliphy80@gmail.com \\
federico.centrone@icfo.eu.}
\affiliation{ICFO-Institut de Ciencies Fotoniques, The Barcelona Institute of Science and Technology, Av. Carl Friedrich Gauss 3, 08860 Castelldefels (Barcelona), Spain.}

\author{Nathan Shettell}
\affiliation{MajuLab, CNRS-UCA-SU-NUS-NTU International Joint Research Laboratory}
\affiliation{Centre for Quantum Technologies, National University of Singapore, 117543 Singapore, Singapore}

\author{Raja Yehia}
\affiliation{ICFO-Institut de Ciencies Fotoniques, The Barcelona Institute of Science and Technology, Av. Carl Friedrich Gauss 3, 08860 Castelldefels (Barcelona), Spain.}

\author{Antonio Acín}
\affiliation{ICFO-Institut de Ciencies Fotoniques, The Barcelona Institute of Science and Technology, Av. Carl Friedrich Gauss 3, 08860 Castelldefels (Barcelona), Spain.}
\affiliation{ICREA - Institució Catalana de Recerca i Estudis Avan\c cats, Lluís Companys 23, 08010 Barcelona, Spain}

\author{Federico Centrone} \thanks{These authors contributed equally and share first authorship. \\ aliphy80@gmail.com \\
federico.centrone@icfo.eu.}
\affiliation{ICFO-Institut de Ciencies Fotoniques, The Barcelona Institute of Science and Technology, Av. Carl Friedrich Gauss 3, 08860 Castelldefels (Barcelona), Spain.}
\affiliation{Universidad de Buenos Aires, Instituto de Física de Buenos Aires (IFIBA), CONICET,
Ciudad Universitaria, 1428 Buenos Aires, Argentina.}

\begin{abstract}
Erasing memory is a fundamental operational task in quantum information processing, governed by Landauer’s principle, which links information loss to thermodynamic work. We introduce and analyze \emph{assisted quantum erasure}, where correlations with a remote system reduce the energetic cost of resetting a memory. We identify \emph{exclusive control} of erasure as the central operational requirement: only a designated party should be able to achieve the minimal cost, whereas any adversary must fail. In the device-dependent regime, we show that entanglement of formation exactly characterizes exclusivity, establishing entanglement as the decisive thermodynamic resource. Moving to a one-sided device-independent scenario, in which only the memory holder’s device is trusted, we develop an operational erasure protocol based on random dephasing and conditional operations. Finally, in a fully device-independent setting, we show how Bell nonlocality and self-testing translate observed violations into lower bounds on any adversary’s erasure capability, yielding a device-independent notion of exclusive thermodynamic control. Taken together, these results elevate quantum erasure from a thermodynamic constraint to an operational primitive: the erasure work cost quantifies secure, exclusive control over quantum memory, ensuring that an unauthorized agent cannot fully erase information under a bounded work budget.
\end{abstract}

\maketitle

\emph{Introduction.}---The deep connection between information and thermodynamics has shaped foundational physics since Maxwell’s demon challenged the second law \cite{maruyama2009colloquium}. Landauer’s principle resolved this paradox by assigning an energetic cost to information erasure: at least $k_B T \ln 2$ of work per erased bit \cite{landauer1961irreversibility, bennett1982thermodynamics, plenio2001physics,Chattopadhyay_2025}. Erasure thus becomes a thermodynamically costly operation, making information a physical resource \cite{auffeves2022quantum}. As quantum processors and networks scale, minimizing this cost becomes essential, especially under repeated operations constrained by finite energy budgets \cite{samsi2023words, li2024unseen}.

In the quantum regime, correlations between systems can be harnessed to reduce the work cost of erasure, allowing one party to lower the entropy of another’s memory using only local operations and a classical message \cite{hotta2011quantum, rodriguez2023experimental, beyer2019steering, biswas2025quantum, ji2022spin}. When this thermodynamic advantage exceeds what is possible with classical correlations, it serves as a witness of nonclassicality. Erasure then becomes more than a resource sink: it is a tool for probing the structure and operational consequences of correlations in distributed quantum systems. Notably, the advantage need not be symmetric, since some parties may enable erasure at a lower cost than others. This raises a fundamental question: can the ability to assist in low-cost erasure be made \emph{exclusive} to a properly correlated party?

We formalize this idea as \emph{exclusive control}: only a designated party, by virtue of their quantum correlations with the system, should be able to assist erasure below a fixed thermodynamic threshold, while any other agent necessarily fails. In this view, erasure is not only a physical process but also a selective form of witnessing, enabled by privileged access to quantum information. Exclusivity is also practically important: in distributed scenarios, one often wants erasure to be enabled only when a designated party cooperates, even if other agents have partial access or side information. This mirrors quantum-cryptographic intuitions, in which only those sharing the appropriate quantum correlations gain secure access, whereas others are fundamentally constrained \cite{branciard2012one, cavalcanti2015detection}. In our setting, untrusted agents may provide limited help, but the full thermodynamic advantage remains inaccessible without genuine quantum correlations.

We study three regimes where exclusivity can be guaranteed. In the {\it device-dependent} setting, where all operations are trusted, exclusivity is determined by entanglement: a trusted helper (Alice) outperforms an adversary (Eve) in assisted erasure precisely when Alice’s entanglement with Bob exceeds Eve’s. In the {\it semi-device-independent} regime, where only Bob’s device is trusted, exclusivity is instead controlled by entropic uncertainty: an operational protocol based on random dephasing and conditional erasure leads to a state-independent lower bound on the assisted work cost that any local-hidden state (LHS) model must obey, and any observed violation of this bound guarantees exclusive control. Finally, in a fully {\it device-independent} scenario, where neither the source nor the measurement devices are characterized, Bell nonlocality and robust self-testing allow observed violations to be converted into lower bounds on the erasure work that any external agent must invest, providing a device-independent notion of exclusive thermodynamic control.

Taken together, these three regimes provide an operational and thermodynamic interpretation of the standard hierarchy of quantum correlations. Entanglement (DD), steering under uncertainty constraints (semi–DI), and Bell nonlocality (DI) each appear as the resource that determines who can make erasure cheap under a given level of trust. The rest of the article makes this hierarchy precise: we formalise assisted erasure and exclusivity, characterise exclusive control in the DD and semi–DI settings, show how to retain recoverability under adversarial attacks, and outline how Bell inequalities and self–testing yield device–independent bounds on exclusive thermodynamic control.

Table~\ref{tab:scenarios} summarizes the three trust scenarios considered in this work.

\begin{table*}[t]

\centering
\small
\begin{tabular}{llll}
\hline\hline
Scenario & Trust assumptions & Quantity & Main statement \\
\hline
DD & Fully trusted state and devices
   & $W_{X\to B}$
   & Theorem~\ref{theoremDDexcl}: exclusivity iff an entanglement-of-formation gap holds. \\
SDI & Bob trusted, Alice untrusted
    & $\widetilde W_{X\to B}$
    & Theorems~\ref{theorem2}--\ref{theorem3}: exclusivity follows from violating the LHS bound. \\
DI & Only Bell statistics trusted
   & $H(Z_B|M_X)$
   & Bell statistics certify a lower bound on any adversary's erasure cost. \\
\hline\hline
\end{tabular}
\caption{Summary of the three trust scenarios considered in this work, with the {\it classical communication channel assumed to be untrusted} in every case. In each scenario, Eve is treated as a competing helper, and the central question is whether she or any other party can match Alice’s reduction of Bob’s erasure cost under the same operational constraints.}
\label{tab:scenarios}
\end{table*}

\begin{figure}
    \centering
\includegraphics[width=0.45\textwidth]{fig1.pdf}
    \caption{Illustration of assisted erasure with competing parties. Bob’s system begins in a high-entropy state, $\rho_B$ (red). Alice and Eve each hold systems correlated with Bob’s, and perform local measurements ($A_M$ and $E_{M}$) to assist in erasure. The resulting conditional states $\rho_{B|A_M}$ and $\rho_{B|E_M}$ show reduced entropy (blue and purple), reflecting Alice’s and Eve’s influence. Although both parties are correlated with Bob, only Alice achieves complete erasure, indicating a stronger correlation than Eve. The thermometer encodes the residual entropy of Bob’s memory, visually distinguishing exclusive thermodynamic control. The figure depicts the direct-erasure realization in which the correlated register $B$ itself is the memory; alternatively, $B$ can serve as the auxiliary reset resource for a separate target memory $M$, as discussed in the text.}
    \label{fig:exclusive-erasure}
\end{figure}

\emph{Assisted Erasure.}---Throughout this work, we consider memories with a degenerate
Hamiltonian. The equilibrium state is therefore maximally mixed and
the reversible work cost is determined entirely by the entropy. Landauer’s principle states that erasing information incurs a fundamental energy cost (and corresponding irreversible heat dissipation) \cite{landauer1961irreversibility, bennett1982thermodynamics, plenio2001physics,Chattopadhyay_2025}: resetting a quantum memory $B$ from $\rho_B$ to a fixed pure state $\ket{0}$ requires, at least,
\begin{equation}
    W_0 = k_B T\, S(B),
\end{equation}
where $S(B) = - \Tr\!\left[\rho_B \ln \rho_B\right]$ is the von Neumann entropy of $B$. Henceforth, we measure entropy in bits and we set $k_B T \ln2 = 1$, so all work costs are measured in units of $k_B T \ln 2$. This baseline applies when the memory is treated in isolation and expresses a fundamental thermodynamic constraint on information processing \cite{landi2021irreversible}.

In practice, erasure need not be performed in isolation. If Bob’s memory $B$ is correlated with a helper system $A$ (by Alice), suitable classical or quantum information-processing can reduce Bob’s \emph{local} work cost \cite{basilewitsch2017beating,esposito2011second, peterson2016experimental, neto2025thermodynamics, beyer2019steering, biswas2025quantum, ji2022spin}. From Bob’s perspective, the work required to reset $B$ is lower, but the gain is fueled by the consumption of the initial $AB$ correlations, so the global thermodynamic accounting remains consistent with Landauer’s principle. In quantitative units, the optimal work bound is governed by the (classical or quantum) conditional entropy,
$W_{\min}=S(B|A)$,
highlighting that pre-established correlations carry a definite thermodynamic value \cite{reeb2014improved,parrondo2015thermodynamics,Huber2015}. Notably, in quantum communication settings, shared entanglement can make the conditional entropy negative, $S(B|A)<0$, so that the corresponding erasure ``cost'' becomes negative and resetting $B$ may even yield net work extraction \cite{rio2011thermodynamic}. In contrast, in the \emph{classical-communication} regime of interest here, where Alice and Bob exchange only classical messages, the (classical) conditional entropy satisfies $S(B|A)\ge 0$, and the erasure cost is therefore necessarily nonnegative.

Formally, Alice performs a measurement $A_M=\{\Pi_k\}$ on her subsystem and announces $k$. Bob then erases conditionally by bringing the system to the state $\ket{0}$. The minimal assisted work cost is
\begin{equation}
W_{A\to B}=\min_{A_{M}} \sum_k p_k\,S(\rho_{B|\Pi_k})
\;=\; S(B)-J_0(B|A),
\end{equation}
with $p_k=\Tr(\rho_A\Pi_k)$ and $\rho_{B|\Pi_k}=\Tr_A[(\Pi_k\otimes I)\rho_{AB}]/p_k$. Here $J_0(B|A)=S(B)-\min_{A_M} S(B|A_M)$ is the standard quantity used to quantify one-way classical correlations (from $A$ to $B$), where the minimum is taken over measurements $A_M$ on $A$ \cite{henderson2001classical}. Hence, any nonzero correlation accessible to Alice lowers Bob’s \emph{local} erasure cost relative to the unassisted baseline $W_0=S(B)$.

Our main focus, however, is not merely cost reduction but \emph{exclusive control} of the erasure: Bob’s memory may also be correlated with an adversary Eve, who could in principle provide side information enabling a similarly low-cost reset. We therefore ask whether one can engineer a regime in which Alice retains \emph{exclusive control}: i.e., she can assist Bob’s erasure within his available work budget, while Eve cannot, despite being correlated with (B). Quantum correlations, owing to their monogamy constraints, are natural candidates to enforce such asymmetric control.

\begin{definition}
    (Exclusive erasure).
Fix a per-run work budget $W_{\max}$ for Bob. We say that Alice holds \emph{exclusive erasure} if
\begin{equation}\label{eq:exclusive-budget}
    W_{A\to B}\ \le\ W_{\max}\ <\ W_{E\to B}\, \leq W_0 .
\end{equation}
The cap must satisfy $W_{\max}<W_0$, otherwise exclusivity is vacuous, since Bob could always erase without any help by paying $W_0$. Fig.~\ref{fig:exclusive-erasure} sketches this setting.
\end{definition} 

Eq.~\eqref{eq:exclusive-budget} is the operational exclusivity we use throughout; the following results guarantee that there \emph{exists} a nontrivial $W_{\max}$ between $W_{A\to B}$ and $W_{E\to B}$. For simplicity, we set Bob's capacity to $W_{\max}=W_{A\to B}$.

The above formulation also admits a useful operational realization in which the correlated register \(B\) is not itself the physical memory containing the information to be erased. Bob may instead hold an independent target memory \(M\), while \(\rho_{AB}\) serves as an assisting thermodynamic resource. In this realization, Alice's measurement and announcement reduce the residual work \(W_{A\to B}\) required to prepare \(B\) in a fixed pure state. Bob can then consume this pure auxiliary state in a trusted, input-independent reset (or irreversible) operation on \(M\), for instance by swapping it with \(M\) and subsequently discarding or rethermalizing the auxiliary system. Eve can analogously attempt to prepare the same reset resource at cost \(W_{E\to B}\). Hence, the exclusivity condition above determines which helper can enable the erasure of the target memory within Bob's available work budget. Importantly, this realization separates the system carrying the assisting correlations from the memory being erased, and allows Bob to test the assisting resource before performing any irreversible operation on the stored information. The formal results below are expressed in terms of the assisted register \(B\), and therefore apply both when \(B\) itself is the memory to be reset and when it is used as the thermodynamic resource for resetting a separate memory \(M\).

In the device-dependent (DD) setting, all systems and measurement devices are fully characterized and trusted. We model any potential adversary, Eve, by assuming that she holds a system $E$, so that $\rho_{AB}=\text{Tr}_E[\rho_{ABE}]$. Classical correlations between Alice and Bob may still lower Bob’s local erasure cost, but they can be freely copied and distributed so that any such advantage can be reproduced by Eve. The following no-go lemma shows that separable states can therefore never generate exclusive erasure.

\begin{lemma}\label{lemma1}
If $\rho_{AB}$ is separable, then any reduction of Bob's erasure cost achieved with Alice's assistance can also be reproduced by Eve. In particular, Alice cannot hold an exclusive erasure, i.e., $W_{A\to B} \geq  W_{E\to B}$.
\end{lemma}

Together with Def.~\eqref{eq:exclusive-budget}, this yields a nontrivial choice of $W_{\max}$, showing that entanglement is a necessary resource for exclusive erasure. For example, if Alice and Bob share a pure entangled state, then $W_{A \rightarrow B}=0$ and, by monogamy, Bob is uncorrelated with any third party. Consequently, $W_{E \rightarrow B}=W_0$ and Alice enjoys strict exclusivity. For mixed states, the situation is subtler: Eve may share correlations with Bob via a purification $\ket{\psi}_{ABE}$. In that case, exclusivity requires that Alice’s correlations with Bob dominate Eve’s across all possible extensions. These correlations can be quantified using the entanglement of formation, an average measure of the entanglement (in the so-called entangled bits, or ebits) required to prepare the state. For a shared bipartite state $\sigma_{AB}=\sum_i p_i \ket{\psi_i}\!\bra{\psi_i}$, the entanglement of formation~\cite{EntangFormat} is defined as
\begin{align*}
    E_f(A\!:\! B)_{\sigma_{AB}}
\coloneqq
\min_{\{p_i,\ket{\psi_i}\}}
\sum_i p_i \,
S\!\left( \operatorname{Tr}_{A}\ket{\psi_i}\!\bra{\psi_i} \right).
\end{align*}

The following theorem makes Alice's exclusive control on Bob's memory explicit in terms of entanglement of formation.

\begin{theorem}\label{theoremDDexcl}
Let Alice and Bob share $\rho_{AB}$ and let $|\psi\rangle_{ABE}$
be a purification, with Eve holding the purifying system. Alice achieves a strictly lower erasure cost than Eve if and only if
\[
E_f(A\!:\!B)_{\rho_{AB}} \;>\;E_f(E\!:\!B)_{\rho_{EB}}.
\]
\end{theorem}

The proof of Lemma~\ref{lemma1} and Theorem~\ref{theoremDDexcl} is provided in the  Supplemental Material. Theorem~\ref{theoremDDexcl} completes the DD characterization: when devices are trusted, exclusive thermodynamic control is governed entirely by an entanglement-of-formation \emph{gap}. Alice can exert exclusive influence if and only if they are more entangled with Bob than any adversary; conversely, separability never suffices (Lemma~\ref{lemma1}). The proof follows from the Koashi–Winter relation \cite{Koashi'2004}, which connects competing classical correlations to competing entanglements of formation and ultimately leads to exclusive erasure.

\emph{Steerability as the Exclusive Resource.}---We now relax trust in Alice’s device and their correlations, while keeping Bob’s operations fully characterized. In this one-sided, semi-device-independent (SDI) regime, Alice’s action is a possibly adversarial black box, so certification cannot rely on its internal form or on entanglement measures. It must come solely from the observed influence of Alice’s announcements on Bob’s system. Throughout, Bob’s local state and operations are characterized and trusted, while the source and Alice’s device are untrusted.

\begin{figure*}
    \centering
    \includegraphics[width=0.99\textwidth]{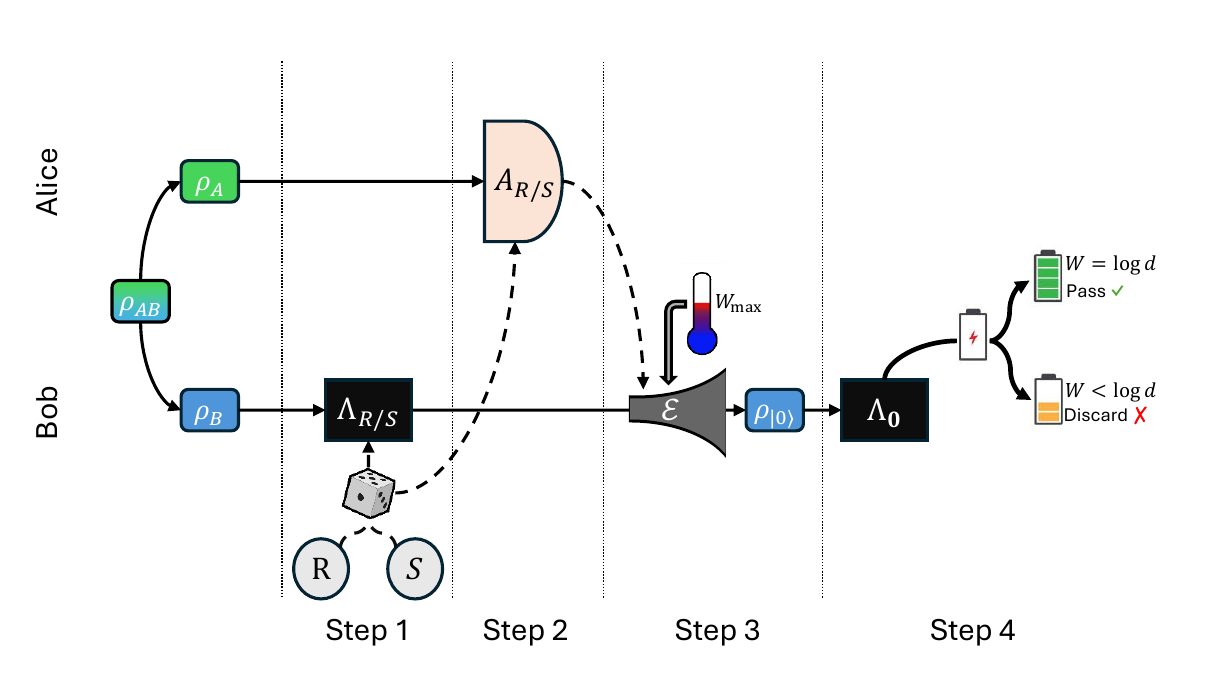}
    \caption{One-sided semi-DI protocol: Bob’s random dephasing, Alice’s announcement, conditional erasure, and verification.}
    \label{fig:SDI-protocol}
\end{figure*}

We adopt a four-step routine that mirrors assisted erasure while enforcing a verifiable benchmark:
\begin{enumerate}
    \item Bob chooses at random one of two orthonormal bases $R=\{\ket{r_i}\}$ or $S=\{\ket{s_j}\}$ announcing the result to Alice. Bob applies a  dephasing channel $\Lambda_R$ or $\Lambda_S$ to $B$, corresponding to the chosen basis. This induces classical distributions $\vec R=\{r_i\}$ and $\vec S=\{s_j\}$ with $r_i=\langle r_i|\rho_B|r_i\rangle$, $s_j=\langle s_j|\rho_B|s_j\rangle$.
    \item Alice performs an (untrusted) measurement on $\rho_A$ and announces a classical outcome. We denote the resulting classical registers by $A_R$ and $A_S$, respectively. 
    \item Bob applies a trusted, outcome-conditioned operation $\mathcal{E}$ to complete the erasure. Bob's budget in this step is limited to $W_{\max}$.
    \item Bob performs a cost-free verification test to certify whether the previous step has indeed erased the state.
\end{enumerate}

The last step is crucial because Bob does not trust Alice’s devices, the shared correlations with memory, or even the classical communication channel; therefore, he must verify that the state has indeed been erased (see Supplemental Material for details).

In general, the random dephasing of the first step of the routine injects entropy into Bob's system, raising the unassisted average erasure cost to
\begin{equation}
    \widetilde W_0=\tfrac12\big(H(R)+H(S)\big)\ \ge\ S(B)=W_0,
\end{equation}
with equality only in special cases (e.g., $\rho_B=\mathbb{I}/d$ or when both the dephasing coincide with the spectral basis of $\rho_B$). Here, $H(X)$ is the Shannon entropy of the probability distribution of the outcomes measured in basis $X$. Crucially, if $R$ and $S$ are incompatible, the Maassen–Uffink relation~\cite{maassen1988generalized} gives a nontrivial \emph{state-independent} bound
\begin{equation}
    \widetilde W_0\ \ge\ \tfrac12\,C(R,S),\qquad
    C(R,S):=\min_{i,j}\log \frac{1}{|\langle r_i|s_j\rangle|^2},
\end{equation}
which depends only on the overlap of the two bases. 

Dephasing in base $R$ yields a classical–quantum state $\sigma^{(R)}_{AR}=\sum_i r_i\,\rho_{A|r_i}\otimes\ket{i}\!\bra{i}_R$. While in Step 2, Alice’s black-box outcome $A_R$ carries at most $I_{\mathrm{acc}}(A_R:R) \leq \chi_R = S\!\left(\sum_i r_i \rho_{A|i}\right) - \sum_i r_i S(\rho_{A|i})$ bits of \emph{accessible} information about $R$ \cite{wilde2013quantum,holevo1973bounds}, implying the classical conditional entropy
\[
H(R|A_R)\ =H(R)- I_{\mathrm{acc}}(A_R:R),
\]
and analogously for base $S$. Optimizing over her black-box choices on each branch gives the operational assisted cost
\begin{equation}
    \widetilde W_{A\to B}=\tfrac12\big(H(R|A_R)+H(S|A_S)\big)\ \le\ \widetilde W_0,
\end{equation}
which depends only on observed announcements and is directly comparable to the state-independent benchmark $\tfrac12\,C(R,S)$.

The random dephasing serves two purposes in the one-sided DI setting: (i) it turns Bob’s quantum state into a classical register ($R$ or $S$) whose uncertainty is quantified by Shannon entropies and constrained by the state-independent complementarity $C(R,S)$; (ii) it yields an operational, device-agnostic metric of help $\widetilde W_{X\to B}$ that depends only on observed announcements, not on trusting Alice’s internal measurements or their initial shared correlations. The dephased classical register in $R$ or $S$ is then reset to $\ket{0}$ in Step 3 by applying an erasure operation chosen according to Alice’s communication (Step~2) and the Bob's dephasing operation (Step~1). 
The physical implementation of this erasure map incurs the associated work cost.

To certify success without reading out the memory, in Step 4 Bob dephases the register in a fixed computational basis and couples it to a Szilard-type work–extraction device, i.e., a thermodynamic engine whose maximal extractable work is achieved only when the register is in a pure state and decreases with any residual entropy.  Perfect erasure leaves a pure state and permits extraction of $\log_2 d$ work per run, while any residual mixing reduces the extractable work (see Supplemental Material).

In the following subsection we leverage this setup: whenever the observed work $\widetilde W_{A\to B}$ drops below the complementarity benchmark $\tfrac12\,C(R,S)$, Alice’s advantage cannot be reproduced by any local hidden state (LHS) model, which in turn underpins \emph{exclusive} control in the one-sided DI setting.
 However, note that, even if the protocol passes verification (step 4) and the observed assisted cost is strictly below the \emph{unassisted} value $\widetilde W_0$, this alone does not certify exclusive control. In particular, if the correlations between Alice and Bob admit an LHS model, the whole procedure can be simulated classically and reproduce the same apparent reduction. In that case, an adversary can match Alice’s help. To capture this limitation, we first bound the assisted cost achievable under LHS models.

\begin{theorem}\label{theorem2}
If the shared state $\rho_{AB}$ admits a local hidden state model (from $A$ to $B$), then
\begin{equation}\label{eq:LHSLHSerasureBound}
    \widetilde W_{A \rightarrow B} \;\geq\; \tfrac12\,C(R,S).
\end{equation}
\end{theorem}

 Intuitively, with an LHS model, Alice’s announcements cannot jointly predict both dephased registers better than complementarity allows \cite{schneeloch2013einstein}; the residual uncertainty imposes a thermodynamic bound on assisted erasure. Using more than two incompatible dephasings on $B$ raises this bound and detects a larger set of steerable states \cite{coles2017entropic}.

In the one-sided DI setting, classical models can again undermine exclusivity. Indeed:

\begin{lemma}%[No-go for LHS states]
If $\rho_{AB}$ admits an LHS model, then one can construct (for any Eve) a strategy with
\begin{equation}
    \widetilde W_{A \rightarrow B} = \widetilde W_{E \rightarrow B}.
\end{equation}
\end{lemma}

\noindent Thus steerability is \emph{necessary} for Alice’s exclusive advantage. It is also \emph{sufficient} once Alice’s observed cost dips below the complementarity benchmark:

\begin{theorem}\label{theorem3}
If
\begin{equation}
    \widetilde W_{A \rightarrow B} \;<\; \tfrac12\,C(R,S),
\end{equation}
then Alice holds an exclusive advantage,
\begin{equation}
    \widetilde W_{A \rightarrow B} \;<\; \widetilde W_{E \rightarrow B}.
\end{equation}
\end{theorem}

\noindent \emph{Proof idea.} The strict violation of the LHS bound in \eqref{eq:LHSLHSerasureBound} certifies steering. Combining this with the entropic uncertainty relations with quantum memory (applied to Eve’s side) yields
$H(R|E)+H(S|E)\!\ge\!C(R,S)$, so any Eve must pay at least $\tfrac12\,C(R,S)$ on average, whereas Alice pays less. Full proofs are provided in the Supplemental Material. These additional noncommuting operations provide a sufficient strategy when neither the shared correlations nor Alice’s measurement device is trusted. Equivalently, any correlations admitting a local-hidden-state (LHS) model necessarily destroy exclusivity, establishing steering as a necessary resource. We show that the additional noncommuting operations are sufficient to guarantee both steering and exclusive erasure. Whether alternative methods of certifying steering can yield a stronger exclusive-erasure advantage in the semi-device-independent setting remains an open question.

In the next section, we show how, when combined with an authentication step that sacrifices some copies of the state, DD and SDI exclusive advantage implies that Eve cannot cause additional irrecoverable loss of 
$AB$ correlations beyond what Bob explicitly chooses to erase.\\

\emph{Memory Retrieval after Adversarial Attacks.}---In both DD and SDI settings, exclusivity implies that a malicious or untrusted helper cannot reproduce the honest assisted cost under the same work cap. In the external-memory realization discussed above, Bob performs the authorization test before acting on the target memory $M$, so a rejected request leaves the stored information untouched. The remaining question is therefore how much of the assisting $AB$ resource must be consumed in order to authenticate the request. In the external-memory realization introduced above, a rejected request leaves the target memory $M$ untouched; the results below instead quantify how much of the assisting $AB$ resource can be preserved and reduce to memory recoverability when $B$ itself is the memory. The following results quantify this preservation in the asymptotic regime.

In what follows, we work in an asymptotic regime consisting of identically and independently distributed (i.i.d.) rounds: Alice and Bob share
$n$ copies of a known state $\rho_{AB}$, Eve holds the purifying system $E$
and may perform arbitrary operations and send classical messages, and Bob's
laboratory and operations
are treated as trusted and noiseless at the level of our idealised analysis.

In the asymptotic i.i.d.\ regime, the relevant resource is the total entropy of the memory state. For a known state $\rho$, each copy carries $S(\rho)$ units of entropy. A Schumacher-type compression \cite{schumacher1995quantum} allows one to apply a global unitary that concentrates all of this entropy essentially into a smaller ``noisy'' subsystem, leaving the remaining registers in a standard pure state. A work budget $W<S(\rho)$ per copy then limits how much of that noisy subsystem can be irreversibly erased; the untouched part can be reversibly re-encoded into fewer intact copies of $\rho$. In Supplemental Material, we formalize this as a compression-based erasure routine.

In our setting, this generic routine is combined with the protocol's \emph{authentication} step, applied only to a random test subset of vanishing relative size. Bob never applies irreversible erasure to the remaining data block unless the compressed test block behaves exactly as predicted for Alice-Bob's honest strategy. Any deviation attributable to Eve is therefore detected during the authentication stage, before the data block is erased. As a result, unauthorized attempts cannot cause additional, undetected loss of Alice–Bob correlations beyond the vanishing test fraction and the part that Bob himself chooses to erase.

It is useful to distinguish the present authentication test from the verification step discussed in the semi-DI protocol. Here Bob does not verify that erasure has already succeeded on the data block. Rather, he uses a sacrificial test subset to check whether the classical control information he receives is consistent with Alice’s honest assistance before committing irreversible work on the remaining copies. In this sense, the subroutine plays the role of an authentication mechanism for the control message. This is relevant because, unlike standard quantum-cryptographic protocols such as QKD, we do not assume from the outset an authenticated classical channel. If such authentication were available in the fully device-dependent case, then Bob would simply ignore any message not certified to come from Alice, and exclusivity would already be enforced at the level of access to the control channel.

\begin{theorem} 
[DD case]\label{thm:dd-recoverability}
Let $\rho_{AB}$ be shared by Alice and Bob with a purification $\ket{\psi}_{ABE}$, and assume the device-dependent exclusivity condition
\[
W_{A\to B} \;<\; W_{E\to B}
\]
holds for some per-run work cap $W_{\max}$ with $W_{A\to B}\leq W_{\max}<W_{E\to B}$. Suppose Bob implements the authentication routine on a random test subset of size $\varepsilon_n n$, where $\varepsilon_n\to0$ and $n\varepsilon_n\to\infty$, and reserves the remaining copies as a data block. Then, under the i.i.d.\ and trust assumptions stated above, for any adversarial strategy by Eve that respects the same work cap, the following holds in the asymptotic limit ($n\to\infty$):
\begin{itemize}
    \item either the authentication step fails, in which case Bob aborts and keeps the untouched data block of size $(1-\varepsilon_n)n+o(n)$,
    \item or the authentication step succeeds, in which case the effective evolution on the data block is asymptotically indistinguishable from Alice’s honest assistance, and Bob can safely erase the memory.
\end{itemize}
In particular, the recoverable fraction satisfies
\[
R_n \;\ge\; 1-\varepsilon_n-o(1)\xrightarrow[n\to\infty]{}1,
\]
and no adversarial strategy can induce additional irrecoverable loss of Alice–Bob correlations beyond Bob’s intentional erasure and the vanishing test fraction.
\end{theorem}

\noindent\emph{Proof idea.} Using the compression routine described above, Bob can reversibly compress the random test block conditioned on Alice’s (or Eve’s) announcement so that, in the honest case, the compressed block has an asymptotically well-defined amount of extractable work, equal to $\varepsilon_n n(1-W_{A\to B})+o(n\varepsilon_n)$. In the dishonest case, even though Bob applies the same honest compression maps, unitarity preserves the total entropy of the compressed test block, so its total extractable work is upper bounded by $\varepsilon_n n(1-W_{E\to B})+o(n\varepsilon_n)$. Since $W_{A\to B}<W_{E\to B}$, these two asymptotic work rates are separated by a nonzero gap, and Bob can choose an acceptance threshold strictly between them. Hence, for sufficiently large $n$, the honest protocol is accepted while any dishonest strategy is rejected. If Bob rejects, only the test subset is consumed and the untouched data block is preserved; if he accepts, the data block can be processed exactly as in the honest protocol under the same work cap.

The same reasoning applies to the semi-device-independent protocol, where the resource is the \emph{post-dephasing} (basis-conditioned) state used in the assisted erasure routine.

\begin{theorem} [Semi-DI case]\label{thm:di-recoverability}
Consider the one-sided device-independent protocol based on random dephasing in bases $R,S$, and let $\tilde{W}_{X\to B}$ denote the observed assisted per-run work cost averaged over the two bases for $X\in\{A,E\}$. Suppose the semi-DI exclusivity condition
\[
\tilde{W}_{A\to B} \;<\; \tilde{W}_{E\to B}
\]
holds for some admissible work cap. If, under the same i.i.d.\ and trust assumptions, Bob applies the authentication strategy on random test subsets of the $R$- and $S$-blocks, of total size $\varepsilon_n n$ with $\varepsilon_n\to0$ and $n\varepsilon_n\to\infty$, then in the asymptotic limit any adversarial attempt either:
\begin{itemize}
    \item fails authentication, in which case Bob aborts and keeps the untouched data block of size $(1-\varepsilon_n)n+o(n)$, or
    \item is asymptotically indistinguishable, on the data block, from Alice’s honest assistance under the same work cap.
\end{itemize}
Consequently, the recoverable fraction again satisfies
\[
R_n \;\ge\; 1-\varepsilon_n-o(1)\xrightarrow[n\to\infty]{}1,
\]
and all \emph{basis-conditioned} $AB$ correlations that fuel the semi-DI protocol remain recoverable up to Bob’s deliberate erasure and the vanishing test fraction.
\end{theorem}

In both regimes, exclusivity therefore comes with a strong robustness guarantee: as long as Bob adheres to the prescribed work cap and verifies only on a vanishing test fraction, no untrusted helper can inflict additional irretrievable damage on his memory beyond that sacrificial fraction. On the remaining data block, either the protocol aborts and the memory is kept untouched, or the post-erasure state is asymptotically the one produced by the honest, Alice-assisted protocol under the same work cap. Any further loss of information arises solely from Bob’s own deliberate erasure.

\emph{Example: two–qubit Werner state.}---
To illustrate the above notions of exclusive control in both the DD and one–sided device–independent settings, we consider the family of two–qubit Werner states
\begin{equation}
    \rho_{AB}(p)
    = p \ketbra{\Psi^-}{\Psi^-} + \frac{1-p}{4}\,\mathds{1}_4,
\end{equation}
with singlet fraction $p\in[0,1]$. Bob’s reduced state is maximally mixed, $\rho_B = \mathds{1}/2$, so the unassisted erasure cost is $W_0 = S(B)=1$.

In the DD regime, Alice’s optimal assisted cost is
\begin{equation}
    W_{A\to B}
    = h_2\!\left(\frac{1-p}{2}\right),
\end{equation}
where $h_2$ is the binary entropy. Eve’s assisted cost is given by the entanglement of formation $E_f(A\!:\!B)$, which for two–qubit Werner states takes the standard Wootters form
\begin{equation}
    W_{E\to B}^{\mathrm{DD}} = E_f(A\!:\!B)
    = h_2\!\left(\frac{1+\sqrt{1-C^2}}{2}\right),
\end{equation}
with concurrence $C = \max\!\left(0,\tfrac{3p-1}{2}\right)$. Exclusive DD control holds whenever $W_{A\to B} < W_{E\to B}^{\mathrm{DD}}$.

In the one–sided semi–DI regime with two mutually unbiased bases on Bob’s side (e.g.\ $\sigma_x$ and $\sigma_z$), Alice’s assisted cost coincides with her DD cost,
\begin{equation}
    \widetilde W_{A\to B} = W_{A\to B},
\end{equation}
while any LHS model must satisfy the state–independent bound
\begin{equation}
    \widetilde W_{E\to B} \;\geq\; \tfrac{1}{2} C(\sigma_x,\sigma_z) = \tfrac{1}{2},
\end{equation}
so semi–DI exclusivity holds whenever $W_{A\to B} < \tfrac{1}{2}$.

\begin{figure}
    \centering
\includegraphics[width=\linewidth]{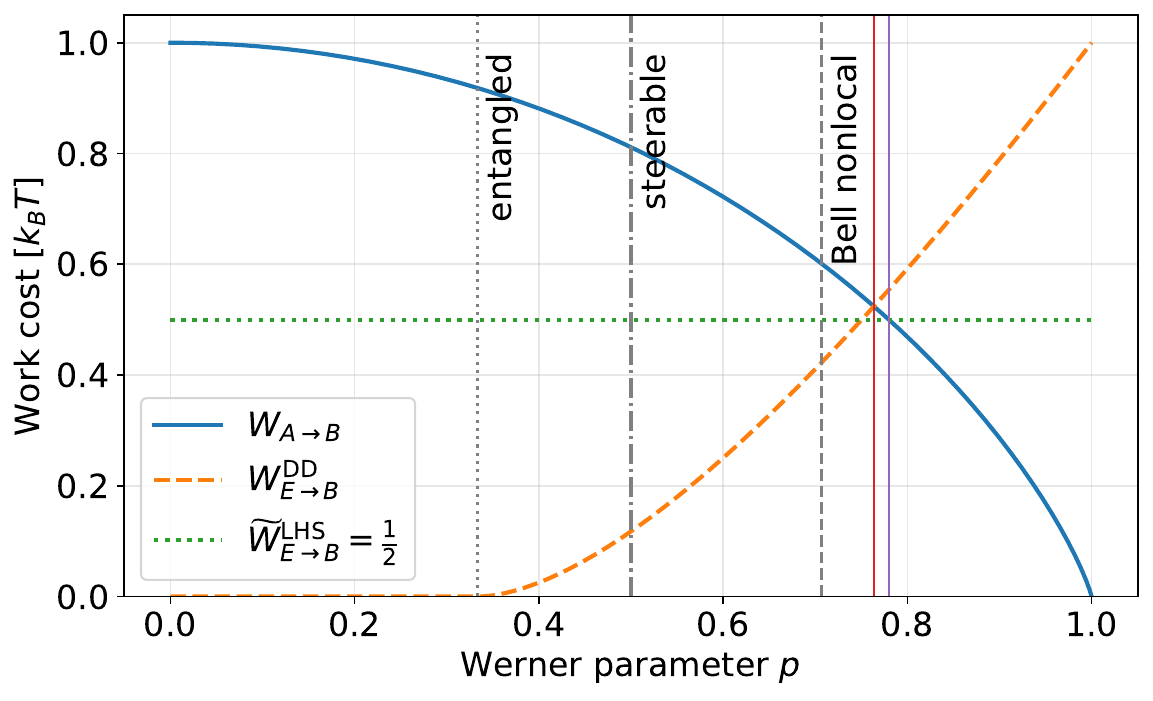}
   \caption{Per–run work costs for assisted erasure on the two–qubit Werner state $\rho_W(p)$.
The solid curve shows Alice's assisted cost in both DD and SDI scenarios $W_{A\to B}= \widetilde W_{A\to B} $, the dashed curve shows Eve's assisted cost 
$W_{E\to B}^{\mathrm{DD}}$ in the device–dependent scenario (given by the entanglement of formation), 
and the dotted horizontal line marks the one–sided device–independent LHS floor 
$\widetilde W_{E\to B}^{\mathrm{LHS}} = \tfrac{1}{2}$ arising from two mutually unbiased dephasings on Bob.
Vertical lines indicate the known thresholds for entanglement ($p>1/3$), steering ($p>1/2$), 
and Bell nonlocality ($p>1/\sqrt{2}$)~\cite{yang2021decompositions}. 
The intersections $W_{A\to B}=W_{E\to B}^{\mathrm{DD}}$ and $W_{A\to B}=\tfrac12$ (not explicitly labelled here) 
mark the onset of exclusive thermodynamic control in the device–dependent and semi–device–independent regimes, witnessing entanglement and steering, respectively.}
    \label{fig:recFrac}
\end{figure}

Figure~\ref{fig:recFrac} shows $W_{A\to B}$, $W_{E\to B}^{\mathrm{DD}}$, and the semi–DI LHS floor $\widetilde W_{E\to B}^{\mathrm{LHS}}=\tfrac{1}{2}$ as functions of the Werner parameter $p$, together with the known thresholds for entanglement, steering, and Bell nonlocality~\cite{yang2021decompositions}. The crossings $W_{A\to B}=W_{E\to B}^{\mathrm{DD}}$ and $W_{A\to B}=\tfrac{1}{2}$ mark, respectively, the onset of exclusive thermodynamic control in the DD and semi–DI regimes. Interestingly, for this highly symmetric family the DD and semi–DI thresholds lie relatively close, indicating that relaxing device assumptions does not dramatically enlarge the parameter range of exclusive control for Werner states. The full derivation of these expressions is given in the Supplemental Material.

\emph{Device-independent thermodynamic control.---} 
We now ask whether exclusive thermodynamic control can be certified without characterizing either the source or the measurement devices. In the fully device-independent (DI) setting, Alice and Bob interact only with black boxes and base their conclusions exclusively on the observed input--output statistics. As usual in a DI protocol, the quantum devices are untrusted, while Bob's classical processing and the thermodynamic apparatus used to store and erase the output register are assumed to operate as prescribed.

Consider a standard Bell scenario in which Alice and Bob choose inputs $x,y$ and obtain outputs $a,b$, defining an observed behaviour $P(a,b|x,y)$. For designated settings, Bob stores his output in a physical classical memory $Z_B$, while $Z_A$ denotes Alice's corresponding classical register. An adversary Eve is allowed to hold an arbitrary quantum system $E$ purifying the underlying realization. Although the state, Hilbert-space dimension, and measurements inside the black boxes are unknown, device-independent methods can infer from the observed behaviour a lower bound
\begin{equation}
H(Z_B|E) \geq h_{\rm DI}[P],
\end{equation}
where $h_{\rm DI}[P]$ is obtained solely from the observed Bell statistics~\cite{acin2007device,pironio2010random}.

This bound has a direct thermodynamic interpretation. Any party assisting Bob is required to communicate with his laboratory through a classical message $M_X$, with $X=A,E$. In our units, the residual work required to erase Bob's physical output memory after receiving such a message is
\begin{equation}
W^{\rm DI}_{X\to B}=H(Z_B|M_X).
\end{equation}
For Eve, the message $M_E$ can be generated by an arbitrary measurement or classical processing of her quantum side information $E$. By data processing,
\begin{equation}
H(Z_B|M_E)\geq H(Z_B|E),
\end{equation}
and therefore
\begin{equation}
W^{\rm DI}_{E\to B}
\geq H(Z_B|E)
\geq h_{\rm DI}[P].
\end{equation}
Hence, for every quantum realization compatible with the observed Bell behaviour, no external party holding the purification can reduce the residual entropy, and therefore the erasure work of Bob's physical memory, below the device-independent value $h_{\rm DI}[P]$. This is the thermodynamic content of the DI certification: the internal physical realization of the black boxes need not be known in order to bound how much an adversary can lower the cost of an irreversible operation in Bob's laboratory.

For the legitimate helper, Alice may simply communicate her observed output, $M_A:=Z_A$. This gives an explicit achievable cost
\begin{equation}
W^{\rm DI}_{A\to B}
\leq H(Z_B|Z_A),
\end{equation}
where $H(Z_B|Z_A)$ is directly determined from the observed Alice--Bob statistics. Consequently, whenever
\begin{equation}
H(Z_B|Z_A)<h_{\rm DI}[P],
\end{equation}
Alice necessarily achieves a strictly lower erasure cost than any adversary,
\begin{equation}
W^{\rm DI}_{A\to B}<W^{\rm DI}_{E\to B}.
\end{equation}
Bob can therefore choose a work threshold satisfying
\begin{equation}
W^{\rm DI}_{A\to B}
\leq W_{\max}
<W^{\rm DI}_{E\to B},
\end{equation}
which certifies exclusive thermodynamic control directly from the observed statistics. Notice that the relevant condition involves both the device-independent lower bound on Eve's residual uncertainty and Alice's observed correlation with Bob; a nonzero Bell violation need not by itself guarantee a positive erasure-cost gap.

For near-maximal Bell violations, self-testing provides a particularly transparent realization of this separation. Robust self-testing results imply that correlations close to the maximal CHSH violation constrain the underlying state and measurements, up to local isometries, to be close to the ideal maximally entangled realization~\cite{mayers2004self,reichardt2013classical,kaniewski2016analytic,supic2020self}. The corresponding Alice--Bob outputs can therefore be made strongly correlated, making $H(Z_B|Z_A)$ small, while the same nonlocal statistics impose a device-independent lower bound on Eve's uncertainty $H(Z_B|E)$. Near the ideal point the two effects combine to yield a finite thermodynamic gap between the work accessible with Alice's assistance and that achievable by any external party.

Importantly, the role of Bell nonlocality here is not merely to certify private information. Through Landauer's principle, the device-independent entropy bound becomes a lower bound on a physical work cost: irrespective of which quantum state and measurements realize the black-box behaviour, Eve cannot assist Bob in erasing the stored output below the certified thermodynamic threshold. Optimizing this bound over the classical messages that may enter Bob's laboratory, or deriving tighter work--Bell trade-offs through semidefinite-program relaxations, could strengthen the quantitative certification. Such refinements are not required, however, for the basic DI conclusion established here: observed nonlocal correlations can certify an exclusive separation in the thermodynamic ability of different parties to assist an irreversible memory-erasure process.

\emph{Discussion.}---We introduced \emph{exclusive control of erasure} as an operational criterion for identifying which parties can lower the erasure cost below a fixed budget. In the device-dependent setting, we established a necessary and sufficient condition: exclusivity arises precisely when the designated helper possesses a strictly larger \emph{entanglement of formation} with Bob than any adversary, implying that classical or separable states cannot yield exclusivity. In the one-sided semi-device-independent regime, exclusivity is instead governed by an \emph{entropic uncertainty threshold}. Random dephasing in two incompatible bases induces a state-independent lower bound $\tfrac12C(R,S)$ on the assisted work cost that any local-hidden-state (LHS) model must obey; only helpers whose announcements beat this bound can achieve exclusive control. At the structural level, the correlations that allow one to surpass the uncertainty threshold are necessarily the steerable ones, but operationally, the relevant parameter is the uncertainty gap set by $C(R,S)$.

More generally, the correlated subsystem need not store the information being erased; it may instead serve as an assisting thermodynamic resource that authorizes the reset of a separate memory, thereby separating the resource that provides exclusive control from the physical register on which the irreversible operation is ultimately performed.

Beyond characterizing when exclusivity holds, we showed that when combined with an authentication step, exclusivity guarantees that malicious or untrusted intervention cannot cause irrecoverable damage beyond a vanishing sacrificial test fraction and Bob’s declared work budget. In both DD and semi-DI settings, the compression routine  is applied only to a random test subset, whose total extractable work is compared with the honest expectation. This allows Bob to treat any external interaction as either (i) a detected failure, in which case he aborts and preserves the untouched data block, or (ii) an effective honest run, in which the data block is asymptotically processed exactly as it would be under Alice’s assistance alone and the same cap $W_{\max}$. Thus, on accepted runs, Eve cannot force additional loss of $AB$ correlations on the data block: Bob never loses more than he \emph{chooses} to erase, apart from the vanishing test fraction used for authentication. This elevates erasure from a passive thermodynamic cost to an active primitive for access control, in which “who can erase” is itself a protected resource.

Our framework naturally admits broader operational connections. Suppose the designated helper is allowed to send quantum, rather than purely classical, messages. In that case, exclusive erasure intersects with the notion of negative conditional entropy \cite{rio2011thermodynamic}, in which quantum side information enables reductions in erasure cost that are unattainable with classical correlations alone. Analyzing exclusivity in this regime could enlarge the class of states that support exclusive thermodynamic control in the device-dependent setting. Moreover, our entropic characterization of the one-sided semi-device-independent protocol resonates with the use of entropic uncertainty relations in quantum cryptography \cite{coles2017entropic}, suggesting that exclusive erasure can serve as a new primitive for resource-constrained cryptographic architectures, in which thermodynamic work bounds act as security witnesses even under limited trust in devices and operations. More broadly, this perspective is consistent with other recent efforts to give steering a direct operational meaning in physical tasks, including metrological and dynamical settings \cite{yadin2021metrological,hsieh2024general,centrone2024breaking}. Taken together, these connections show that the hierarchy of quantum correlations is also a hierarchy of exclusive thermodynamic control over distant memories.

See the Supplemental Material for detailed proofs and additional derivations.

\begin{acknowledgments}
\textbf{Acknowledgement}. This work was supported by the Government of Spain (Severo Ochoa CEX2019-000910-S, FUNQIP and NextGeneration EU PRTR-C17.I1) and  European Union (QSNP, 101114043, Quantera Veriqtas), Fundació Cellex, Fundació Mir-Puig, the Government of Spain (Severo Ochoa CEX2019-000910-S and FUNQIP), Generalitat de Catalunya (CERCA program). M.A. acknowledges funding by the European Union  (QURES, 101153001). N.S. is supported by the National Research Foundation, Singapore, through the National Quantum Office, hosted in A*STAR, under its Centre for Quantum Technologies Funding Initiative (S24Q2d0009). In addition, N.S. acknowledges the Plan France 2030 through the project BACQ as part of the MetriQs-France program (Grant ANR-22-QMET-0002). R.Y. acknowledges funding from the Horizon Europe program (QUCATS) as well as LUBOS. A.A. acknowledges the ERC AdG CERQUTE, the AXA Chair in Quantum Information Science. F.C. acknowledges funding by the European Union (EQC, 101149233). Views and opinions expressed are, however, those of the authors only and do not necessarily reflect those of the European Union or European Research Executive Agency. Neither the European Union nor the granting authority can be held responsible for them.
\end{acknowledgments}

{\bf Data availability statement.} No datasets were generated or analyzed in this study. All numerical results and figures were obtained directly from the analytical expressions presented in the manuscript.

\vspace{5pt}
% \clearpage
\twocolumngrid
%\bibliography{biblio}

%

\end{thebibliography}

\newpage
\onecolumngrid

\appendix

We use the notation and definitions introduced in the main text throughout.

\section{Proof of Lemma 1} 

Before proving the theorem, we can prove lemma 1. It is a specific case of the above theorem, but sets the stage for the formalism we will employ.

\begin{proof}
A classical ensemble admits a decomposition
\begin{equation}
    \rho_{AB} = \sum_i p_i \, \dyad{\psi_i}_A \otimes \dyad{\phi_i}_B.
\end{equation}
Since the state is separable, we may straightforwardly make the following tripartite extension
\begin{equation}
    \rho_{ABE} = \sum_i p_i \, \dyad{\psi_i}_A \otimes \dyad{\phi_i}_B \otimes \dyad{\psi_i}_E,
\end{equation}
for which $\rho_{AB} = \rho_{AE}$. This shows that Eve can replicate Alice’s correlations with Bob and achieve the same erasure cost. Moreover, Eve can assign a perfectly distinguishable register that encodes the classical index $i$, e.g., 
\begin{equation}
    \sigma_{ABE} = \sum_i p_i \, \dyad{\psi_i}_A \otimes \dyad{\phi_i}_B \otimes \dyad{i}_E,
\end{equation}
where $\braket{i|j}=\delta_{i,j}$. In this case, $J_o(B|E) = S(B) \geq J_o(B|A)$, which implies
\begin{equation}
    W_{A \rightarrow B} \geq W_{E \rightarrow B}.
\end{equation}
Thus, classical correlations alone never grant Alice exclusive control over Bob’s memory.
\end{proof}

\section{Proof of Theorem 1}

Now we show the proof of Theorem 1.

\begin{proof}
Consider the cost gap
\begin{align}
    \Delta &= W_{E \rightarrow B} - W_{A \rightarrow B} \nonumber \\
           &= J_o(B|A) - J_o(B|E).
\end{align}
The Koashi--Winter relation~\cite{Koashi'2004} states that
\begin{equation}
    J_o(B|A) + E_f(E:B) = J_o(B|E) + E_f(A:B) = S(B).
\end{equation}
Substituting, we find
\begin{equation}
    \Delta = E_f(A:B) - E_f(E:B).
\end{equation}
Hence, Alice achieves exclusive advantage, i.e., $\Delta >0$, if and only if $E_f(A:B) > E_f(E:B)$, independent of the purification.
\end{proof}

\section{Verification step in the SDI setting}

In the one-sided semi-device-independent setting, Bob's operations are trusted, but neither the
shared state nor Alice's measurement device are assumed to be reliable. As a result, even if
Alice's announcement lowers Bob's \emph{average} assisted erasure cost, it does not by itself
guarantee that Bob's memory has actually been reset to the target pure state in each run.
A verification step is therefore necessary: without it, Bob could apply the prescribed correction
and still end up with a mixed state, while incorrectly concluding that erasure has succeeded.

A simple example illustrates the point. Suppose Alice and Bob are meant to share the Bell state
\[
\ket{\phi^+}=\frac{\ket{00}+\ket{11}}{\sqrt 2},
\]
and Bob follows the honest protocol based on Alice's announced basis and outcome. If the shared
state and Alice's measurements are indeed the intended ones, Bob's conditional correction resets
his memory to $\ket{0}$ at zero work cost. However, if Alice's device measures in the wrong basis,
or if the shared state differs from the expected one, then the same announced message may leave
Bob's conditional state mixed. In that case the prescribed unitary correction no longer produces
the erased state, even though the observed assisted cost may still satisfy
\(\widetilde W_{A\to B}<\widetilde W_0\).

For this reason, in the semi-DI protocol Bob must verify that the reset has actually succeeded.
He does so by testing the final memory after the erasure step. In the honest case, the memory is in
the pure state $\ket{0}$ and Bob can extract the maximal work \(\log d\) from it using an
ideal work extractor. Any shortfall certifies that the reset was incomplete, either because the
shared state was not the intended one, or Alice's device was faulty. In such cases, the protocol must be aborted (or Bob must pay
additional work to complete the erasure).

\section*{Proof of Theorem 2}
\begin{proof}
Assume that $\rho_{AB}$ admits a local--hidden--state (LHS) model from $A$ to $B$.
Then there exist a classical variable $\lambda$ with distribution $\{p_\lambda\}$,
a family of states $\{\rho_B^\lambda\}$ on Bob's system, and classical response
probabilities $p(a|x,\lambda)$ such that, for every measurement setting $x$ on $A$
and outcome $a$, Bob's (unnormalized) conditional state can be written as
\begin{equation}\label{eq:LHSmodel}
\sigma_{a|x}\;:=\;\Tr_A\!\big[(M_{a|x}\otimes \mathbbm{1})\,\rho_{AB}\big]
\;=\;\sum_\lambda p_\lambda\,p(a|x,\lambda)\,\rho_B^\lambda.
\end{equation}
In words: before any measurement choice is made, Bob's system is drawn from a
pre-existing ensemble $\{\rho_B^\lambda\}$, while Alice's announced outcome is
generated by classical post-processing of $\lambda$.

We now consider the semi-DI protocol in the main text, in which Bob chooses uniformly
at random one of two dephasing settings $R$ or $S$ (with orthonormal bases
$R=\{\ket{r_i}\}$ and $S=\{\ket{s_j}\}$), applies the corresponding dephasing to $B$,
and Alice announces a classical message (denoted $A_R$ or $A_S$ depending on the setting).
The semi-DI assisted work cost is defined operationally as
\begin{equation}\label{eq:Wtilde_def}
\widetilde W_{A\to B}\;=\;\frac12\Big(H(R|A_R)+H(S|A_S)\Big),
\end{equation}
where $H(\cdot|\cdot)$ are classical conditional Shannon entropies of Bob's dephasing
outcome given Alice's announcement.

Fix $\lambda$. If Bob dephases in basis $R$, then the outcome distribution is
\[
p_i^{(R,\lambda)}:=\bra{r_i}\rho_B^\lambda\ket{r_i},
\qquad
\text{and similarly}\qquad
p_j^{(S,\lambda)}:=\bra{s_j}\rho_B^\lambda\ket{s_j}.
\]
Since Alice's announcement is obtained from $\lambda$ by classical post-processing
(and does not give access to anything beyond $\lambda$), it cannot reduce Bob's
uncertainty below what would be possible if $\lambda$ itself were known. Hence
\begin{equation}\label{eq:cond_lower}
H(R|A_R)\;\ge\;\sum_\lambda p_\lambda\,H\!\big(\vec p^{\,(R,\lambda)}\big),
\qquad
H(S|A_S)\;\ge\;\sum_\lambda p_\lambda\,H\!\big(\vec p^{\,(S,\lambda)}\big),
\end{equation}
where $\vec p^{\,(R,\lambda)}=\{p_i^{(R,\lambda)}\}_i$ and $\vec p^{\,(S,\lambda)}=\{p_j^{(S,\lambda)}\}_j$.
Substituting \eqref{eq:cond_lower} into \eqref{eq:Wtilde_def} gives
\begin{equation}\label{eq:Wtilde_lambda}
\widetilde W_{A\to B}
\;\ge\;\frac12\sum_\lambda p_\lambda\Big(
H\!\big(\vec p^{\,(R,\lambda)}\big)+H\!\big(\vec p^{\,(S,\lambda)}\big)\Big).
\end{equation}

Now, for each fixed $\lambda$, the two distributions $\vec p^{\,(R,\lambda)}$ and
$\vec p^{\,(S,\lambda)}$ arise from measuring the \emph{same} state $\rho_B^\lambda$
in the two bases $R$ and $S$. Therefore, the Maassen--Uffink entropic uncertainty
relation yields
\begin{equation}\label{eq:MU}
H\!\big(\vec p^{\,(R,\lambda)}\big)+H\!\big(\vec p^{\,(S,\lambda)}\big)
\;\ge\; C(R,S),
\qquad
C(R,S):=\min_{i,j}\log\frac{1}{|\braket{r_i|s_j}|^2}.
\end{equation}
Averaging \eqref{eq:MU} over $\lambda$ and using \eqref{eq:Wtilde_lambda} concludes that
\[
\widetilde W_{A\to B}\;\ge\;\frac12\,C(R,S),
\]
which is exactly (7).  This proves the theorem.
\end{proof}

\section*{Proof of Lemma~2}
\begin{proof}
Assume that $\rho_{AB}$ admits a local--hidden--state (LHS) model from $A$ to $B$.
Then there exist a random variable $\lambda$ with distribution $\{p_\lambda\}_\lambda$,
a collection of states $\{\sigma_B^\lambda\}_\lambda$, and response functions
$q_A(k|X,\lambda)$ (for $X\in\{R,S\}$) such that the (unnormalised) conditional states on
Bob satisfy
\begin{equation}
\label{eq:LHSmodel-lemma2}
\sigma_{B|k,X}
=\sum_{\lambda} p_\lambda\, q_A(k|X,\lambda)\,\sigma_B^\lambda .
\end{equation}
In particular, after Bob applies the dephasing $\Lambda_X$ and regards $B$ as the
classical register $X_B$, the induced joint distribution of $(X_B,A_X)$ is
\[
P(i,k|X)=\sum_{\lambda} p_\lambda\, q_A(k|X,\lambda)\,
\bra{x_i}\sigma_B^\lambda\ket{x_i},
\]
where $\{\ket{x_i}\}$ is the basis defining $\Lambda_X$.

Now consider the tripartite extension in which Eve holds a classical register storing
$\lambda$. Upon learning $X$, Eve outputs a message $E_X$ by sampling
$q_E(k|X,\lambda):=q_A(k|X,\lambda)$. Then the induced distribution of $(X_B,E_X)$ is
\emph{identical} to that of $(X_B,A_X)$ for each $X\in\{R,S\}$, hence
$H(X_B|E_X)=H(X_B|A_X)$ for both settings. Averaging over $R,S$ gives
\[
\widetilde W_{E\to B}
=\tfrac12\big(H(R|E_R)+H(S|E_S)\big)
=\tfrac12\big(H(R|A_R)+H(S|A_S)\big)
=\widetilde W_{A\to B}.
\]
This constructs an Eve strategy matching Alice’s semi-DI assisted cost, proving the lemma.
\end{proof}

\section{Proof of Theorem 3}

\begin{proof}
Let
\[
\Sigma_A := H(R|R')+H(S|S'),
\qquad
\Sigma_E := H(R|R'')+H(S|S''),
\]
so that
\[
\widetilde W_{A\to B}=\frac{\Sigma_A}{2}\,,
\qquad
\widetilde W_{E\to B}=\frac{\Sigma_E}{2}\,.
\]

The hypothesis of the theorem,
\[
\widetilde W_{A\to B}<\frac{C(R,S)}{2}\,,
\]
is equivalent to
\begin{equation}
\Sigma_A < C(R,S).
\label{eq:SigmaA_bound}
\end{equation}

Now let $E$ denote Eve's quantum side information before she measures it to generate the
classical outcomes $R''$ and $S''$. Since $R''$ and $S''$ are obtained from $E$ by measurement
and classical post-processing, data processing implies
\begin{equation}
H(R|R'') \ge H(R|E),
\qquad
H(S|S'') \ge H(S|E).
\label{eq:data_processing_E}
\end{equation}
Similarly, since $R'$ and $S'$ are classical outcomes obtained by measuring Alice's quantum
system $A$,
\begin{equation}
H(R|R') \ge H(R|A),
\qquad
H(S|S') \ge H(S|A).
\label{eq:data_processing_A}
\end{equation}

For any tripartite extension $\rho_{ABE}$, Berta \emph{et al.}~\cite{berta2010uncertainty}
proved the entropic uncertainty relations with quantum memory
\begin{equation}
H(R|E)+H(S|A)\ge C(R,S),
\label{eq:Berta1}
\end{equation}
and
\begin{equation}
H(S|E)+H(R|A)\ge C(R,S).
\label{eq:Berta2}
\end{equation}
Adding \eqref{eq:Berta1} and \eqref{eq:Berta2} gives
\[
H(R|E)+H(S|E)+H(R|A)+H(S|A)\ge 2C(R,S).
\]
Using \eqref{eq:data_processing_A}, we obtain
\[
H(R|E)+H(S|E)
\ge 2C(R,S)-\Sigma_A.
\]
Finally, by \eqref{eq:data_processing_E},
\[
\Sigma_E
=H(R|R'')+H(S|S'')
\ge H(R|E)+H(S|E)
\ge 2C(R,S)-\Sigma_A.
\]
Since \(\Sigma_A<C(R,S)\) by \eqref{eq:SigmaA_bound}, it follows that
\[
\Sigma_E > \Sigma_A.
\]
Multiplying by \(1/2\) yields
\[
\widetilde W_{E\to B}>\widetilde W_{A\to B},
\]
which proves the theorem.
\end{proof}

\section{Derivation details for the Werner state example}

We now illustrate our framework using the two–qubit Werner state~\cite{werner1989quantum}
\begin{align}
    \rho_{AB}(p)
    &= p \ketbra{\Psi^-}{\Psi^-}+\frac{1-p}{4}\mathds{1}_4 \\
    &= \frac{1}{4}\left( \mathds{1}_A\otimes\mathds{1}_B 
    - p\sum_{i=x,y,z} \sigma_{i_A}\otimes \sigma_{i_B}\right),
\end{align}
where $\ket{\Psi^-}$ is the singlet state and $\sigma_i$ are the Pauli matrices.  
The Pauli parametrization is particularly convenient for writing the conditional states after local measurements~\cite{lyons2011entanglement, li2021information}.

The unconditional local state of Bob is maximally mixed,
\[
\rho_B=\Tr_A[\rho_{AB}]=\frac{\mathds{1}}{2},
\]
so its stand–alone erasure cost is $W_0=S(B)=1$ (in $k_BT$ units).

The conditional local state after Alice’s projective measurement and announcement is
\begin{equation}
    \rho^{\pm}_{B|A}
    =\Tr_A\big[(\Pi^{\pm}_A\otimes\mathds{1}_B)\,\rho_{AB}\big],
\end{equation}
where $\Pi^{\pm}_A=\frac{1}{2}\big(\mathds{1}\pm\sum_i r_i \sigma_i\big)$ defines Alice’s local measurement along Bloch direction $\vec r=(r_x,r_y,r_z)$, and the sign corresponds to the $\pm$ outcome. For measurements along a Pauli axis, one finds explicitly
\begin{equation}
    \rho^{\pm}_{B|A}
    =\frac{1}{2}
    \begin{pmatrix}
        1\mp p & 0 \\[2pt]
        0 & 1\pm p
    \end{pmatrix}.
\end{equation}

The assisted work cost is
\begin{gather}\label{eq:WA_def}
    W_{A\to B} \;=\; S(B)-J(B|A) \\
    = 1-\big[2-S(\rho_{AB})-D(B|A)\big]
    = S(\rho_{AB})+D(B|A)-1,
\end{gather}
where we used $S(B)=1$ for two–qubit Werner states and the identity
$J(B|A)=2-S(\rho_{AB})-D(B|A)$ between one–way classical correlation and discord.

For the Werner family,
\[
S(\rho_{AB})
= -\,\frac{1+3p}{4}\,\log_2\!\frac{1+3p}{4}
   \;-\;3\,\frac{1-p}{4}\,\log_2\!\frac{1-p}{4},
\]
\[
D(B|A)
= \frac{1-p}{4}\log_2(1-p)
  -\frac{1+p}{2}\log_2(1+p)
  +\frac{1+3p}{4}\log_2(1+3p).
\]
Inserting these into~\eqref{eq:WA_def} and simplifying yields
\begin{gather}
   W_{A\to B}
   = 1 - \frac{1-p}{2}\,\log_2(1-p) 
       - \frac{1+p}{2}\,\log_2(1+p) \\
   = h_2\!\left(\frac{1-p}{2}\right),
\end{gather}
where $h_2(x)=-x\log_2 x-(1-x)\log_2(1-x)$ is the binary entropy.

To compute the work $W_{E\to B}$ associated with an adversarial helper in the DD scenario, we use that in our setting it is given by the entanglement of formation of the Werner state~\cite{diaz2018exact},
\begin{equation}
   W_{E\to B} \;=\; E_f(A\!:\!B),
\end{equation}
with
\begin{equation}
   E_f(A\!:\!B)
   = h_2\!\left( \frac{1+\sqrt{1-C^2}}{2} \right),
\end{equation}
where the concurrence is $C=\max\!\left(0,\tfrac{3p-1}{2}\right)$.

In the semi–DI scenario we dephase Bob in two mutually unbiased bases, say the eigenbases of $\sigma_z$ and $\sigma_x$, and define
\begin{equation}
    \widetilde{W}_{A\to B}
    =\frac{1}{2}\Big(H(\sigma_{z_B}|\sigma_{z_A})
                    +H(\sigma_{x_B}|\sigma_{x_A}) \Big),
\end{equation}
where $H(\cdot|\cdot)$ are classical conditional entropies of the joint outcome distributions.  
For the Werner state, the error probability in any Pauli basis is $(1-p)/2$, so each contribution equals $h_2\big(\tfrac{1-p}{2}\big)$, and hence
\begin{equation}
    \widetilde{W}_{A\to B} = W_{A\to B}.
\end{equation}
On the other hand, Eve’s assisted erasure cost in the one–sided DI setting is bounded below by the LHS floor from Theorem 2,
\begin{equation}
    \widetilde{W}_{E\to B}\;\geq\; \frac{1}{2}C(\sigma_z,\sigma_x)
    =\frac{1}{2}\log_2 2 = \frac{1}{2}.
\end{equation}

Putting these expressions together, the per–run work costs $W_{A\to B}$, $W_{E\to B}^{\mathrm{DD}}=E_f$, and the semi–DI LHS bound $\widetilde W_{E\to B}^{\mathrm{LHS}}=\tfrac12$ as functions of the Werner parameter $p$ are shown in Fig. 3.

\section*{Memory Erasure with bounded energy}
\label{app:rev-erasure}

In this section, we recall a standard state-transformation routine that we use as a building block in our recoverability arguments. It captures the idea that, for a \emph{known} memory state and a bounded work budget, one can reversibly compress the entropy into a smaller subsystem, erase part of it, and then recover a reduced number of intact copies of the original state.

\begin{lemma}
\label{lem:rev-erasure}
    Consider $n$ copies of a quantum memory state $\rho \in D(\mathcal{H}_d)$ that we wish to erase, where the available work per copy is bounded by $W \leq S(\rho)$. Then, in the asymptotic limit $n \to \infty$, there exists a protocol consisting of
    \begin{itemize}
        \item unitaries $U_1$ and $U_2$ (thermodynamically reversible), and
        \item an irreversible erasure channel $\Lambda_{\mathrm{irr}}$ applied to a subset of registers,
    \end{itemize}
    such that the overall map
    \[
        \mathcal{E} = U_2 \circ \Lambda_{\mathrm{irr}} \circ U_1
    \]
    implements
    \[
        \rho^{\otimes n}\;\xrightarrow{\ \mathcal{E}\ }\; \rho^{\otimes n(1-W/S(\rho))}\!,
    \]
    i.e., the protocol irreversibly removes exactly a fraction $W/S(\rho)$ of the total entropy and leaves a fraction
    \[
        R \;=\; 1 - \frac{W}{S(\rho)}
    \]
    of the memory intact, in the form of $(1-W/S(\rho))\,n$ effective copies of $\rho$.
\end{lemma}

\begin{proof}
    Suppose Bob holds $n$ copies of $\rho \in D(\mathcal{H}_d)$. The total entropy is $n S(\rho)$, and the total work budget is $nW$.

    \medskip\noindent\textbf{Step 1.}
    By the standard Shannon--Schumacher compression picture, there exists a unitary $U_1$ (possibly acting jointly on the $n$ systems and an ancilla) such that
    \[
        \rho^{\otimes n}
        \;\xrightarrow{U_1}\;
        \ket{0}\!\bra{0}^{\otimes n\bigl(1-\frac{S(\rho)}{\log d}\bigr)}
        \;\otimes\;
        \Bigl(\tfrac{I}{d}\Bigr)^{\otimes \frac{n S(\rho)}{\log d}}.
    \]
    The first factor is a large pure register with no entropy, and the second factor is a tensor product of maximally mixed $d$-dimensional registers containing essentially all the entropy. Since $U_1$ is unitary, this step is reversible and costs no work.

    \medskip\noindent\textbf{Step 2.}
    Erasing one $d$-dimensional maximally mixed register costs $\log d$ units of work. With total budget $nW$, Bob can therefore erase
    \[
        \frac{nW}{\log d}
    \]
    such registers. Define the irreversible erasure channel
    \[
        \Lambda_{\mathrm{irr}}(\sigma) = \ket{0}\!\bra{0}
        \quad \forall\,\sigma \in D(\mathcal{H}_d),
    \]
    and apply it to exactly $\frac{nW}{\log d}$ of the maximally mixed registers:
    \[
        \Bigl(\tfrac{I}{d}\Bigr)^{\otimes \frac{n S(\rho)}{\log d}}
        \;\xrightarrow{\Lambda_{\mathrm{irr}}^{\otimes \frac{nW}{\log d}}}\;
        \ket{0}\!\bra{0}^{\otimes \frac{n W}{\log d}}
        \;\otimes\;
        \Bigl(\tfrac{I}{d}\Bigr)^{\otimes \frac{n S(\rho)}{\log d}\left(1-\frac{W}{S(\rho)}\right)} .
    \]
    This consumes exactly $nW$ units of work.
    
\medskip\noindent\textbf{Step 3 (Reversible recovery of fewer copies).}
    The remaining maximally mixed registers carry a total entropy of
    \[
        n S(\rho)\Bigl(1-\frac{W}{S(\rho)}\Bigr) = n\bigl(S(\rho)-W\bigr).
    \]
    By Schumacher dilution, there exists a unitary $U_2$ such that
    \[
        \ket{0}\!\bra{0}^{\otimes *}
        \otimes
        \Bigl(\tfrac{I}{d}\Bigr)^{\otimes \frac{n S(\rho)}{\log d}\left(1-\frac{W}{S(\rho)}\right)}
        \;\xrightarrow{U_2}\;
        \rho^{\otimes n(1-W/S(\rho))},
    \]
    where the precise number of $\ket{0}$-registers is irrelevant. Again, $U_2$ is reversible and cost-free.

    \medskip

    Composing the three steps, the overall transformation is
    \[
        \rho^{\otimes n}
        \;\xrightarrow{\ \mathcal{E}=U_2 \circ \Lambda_{\mathrm{irr}} \circ U_1\ }\;
        \rho^{\otimes n(1-W/S(\rho))},
    \]
    so a fraction $W/S(\rho)$ of the total entropy is erased and a fraction $R=1-W/S(\rho)$ of the memory remains in intact copies of $\rho$.
    
\end{proof}

\section{Proof of Theorem 4}

\begin{proof}

Suppose Bob receives \(n\) copies of a memory state. In the honest case, Alice's classical communication induces an ensemble
\[
\{p_i^A,\rho_i^A\}_{i=1}^N,
\qquad
p_i^A=\frac{n_i^A}{n},
\]
and the average work required to erase Bob's memory is
\[
W_{A\to B}=\sum_i p_i^A S(\rho_i^A).
\]
In the dishonest case, Eve may instead induce an unknown ensemble
\[
\{p_i^E,\rho_i^E\}_{i=1}^N,
\qquad
p_i^E=\frac{n_i^E}{n},
\]
with corresponding cost
\[
W_{E\to B}=\sum_i p_i^E S(\rho_i^E).
\]
We assume the exclusivity condition
\[
W_{A\to B} < W_{E\to B},
\]
and define the gap
\[
\Delta := W_{E\to B}-W_{A\to B} > 0.
\]

Let \(\varepsilon_n\in(0,1)\) be such that
\[
\varepsilon_n\to0,
\qquad
n\varepsilon_n\to\infty
\qquad
(n\to\infty).
\]
For each label \(i\), Bob sets aside a test block of size
\[
m_i^X := \lfloor \varepsilon_n n_i^X \rfloor,
\qquad X\in\{A,E\},
\]
and leaves the remaining \((n_i^X-m_i^X)\) copies untouched. The untouched part therefore has total size
\[
n_{\rm data}=n-\sum_i m_i^X = (1-\varepsilon_n)n + o(n).
\]

For each \(i\), Bob applies to the test copies the Schumacher compression unitary \(U_i\) designed for the honest state \(\rho_i^A\). By Lemma~\ref{lem:rev-erasure}, in the honest i.i.d.\ case,
\begin{equation}\label{eq:honest-compression-proof}
(\rho_i^A)^{\otimes m_i^A}
\;\xrightarrow{U_i}\;
(\ket{0}\bra{0})^{\otimes m_i^A(1-S(\rho_i^A))}
\otimes \text{(noisy block)},
\end{equation}
up to vanishing typicality errors. Summing over all \(i\), the compressed test block contains
\[
K_A
=
\sum_i m_i^A \bigl(1-S(\rho_i^A)\bigr)
=
\varepsilon_n n \bigl(1-W_{A\to B}\bigr)+o(n\varepsilon_n)
\]
pure qudits and a noisy block carrying the remaining entropy.

Bob now couples the \emph{entire} compressed test block to an ideal work extractor. Since only the pure part contributes work, the total extractable work in the honest case is
\begin{equation}\label{eq:WextA}
W_{\rm ext}^{(A)}
=
K_A + o(n\varepsilon_n)
=
\varepsilon_n n\bigl(1-W_{A\to B}\bigr)+o(n\varepsilon_n).
\end{equation}

In the dishonest case, Bob applies the same honest compression unitaries \(U_i\) to the test copies \((\rho_i^E)^{\otimes m_i^E}\). The output need not have the form \eqref{eq:honest-compression-proof}, but the total entropy of the compressed test block is preserved because the map is unitary:
\[
S\!\left(U_i(\rho_i^E)^{\otimes m_i^E}U_i^\dagger\right)
=
m_i^E S(\rho_i^E).
\]
Hence the total entropy of the full compressed dishonest test block is
\[
S(\Omega_E^{\rm test})
=
\sum_i m_i^E S(\rho_i^E)
=
\varepsilon_n n\,W_{E\to B}+o(n\varepsilon_n).
\]
For a block of \(m\) qudits in state \(\omega\), the maximal extractable work is \(m-S(\omega)\) (in units of \(k_BT\ln2\)). Therefore the total extractable work from the dishonest test block is bounded by
\begin{equation}\label{eq:WextE}
W_{\rm ext}^{(E)}
\le
\varepsilon_n n\bigl(1-W_{E\to B}\bigr)+o(n\varepsilon_n).
\end{equation}

By \eqref{eq:WextA} and \eqref{eq:WextE}, the honest and dishonest asymptotic work yields are
\[
W_{\rm ext}^{(A)}
=
\varepsilon_n n\bigl(1-W_{A\to B}\bigr)+o(n\varepsilon_n),
\qquad
W_{\rm ext}^{(E)}
\le
\varepsilon_n n\bigl(1-W_{E\to B}\bigr)+o(n\varepsilon_n).
\]
Since \(W_{A\to B}<W_{E\to B}\), the two leading-order work rates are separated by the gap
\[
\Delta:=W_{E\to B}-W_{A\to B}>0.
\]

Set
\[
W_{\rm th}:=\varepsilon_n n\left(1-\frac{W_{A\to B}+W_{E\to B}}{2}\right),
\]
namely the midpoint between the honest and dishonest asymptotic work values. Then
\[
W_{\rm ext}^{(A)}-W_{\rm th}
=
\frac{\Delta}{2}\,\varepsilon_n n+o(n\varepsilon_n),
\]
whereas
\[
W_{\rm ext}^{(E)}-W_{\rm th}
\le
-\frac{\Delta}{2}\,\varepsilon_n n+o(n\varepsilon_n).
\]
Because \(n\varepsilon_n\to\infty\), the \(o(n\varepsilon_n)\) corrections are negligible compared
with \(\varepsilon_n n\Delta\). Therefore, for all sufficiently large \(n\),
\[
W_{\rm ext}^{(A)}>W_{\rm th},
\qquad
W_{\rm ext}^{(E)}<W_{\rm th}.
\]

Bob can thus use the deterministic block-work test: he accepts iff the extracted work from the
compressed test block satisfies
\begin{equation}\label{eq:acceptance-DD}
W_{\rm ext}\ge W_{\rm th}.
\end{equation}
 
In the honest i.i.d.\ case, the Schumacher asymptotics imply that Bob accepts with probability \(1-o(1)\). In the dishonest case, \eqref{eq:WextE} implies that Eve cannot reach the threshold \eqref{eq:acceptance-DD} for sufficiently large \(n\), so the acceptance probability tends to zero.

If Bob rejects, he discards only the test block and keeps the untouched data block of size
\[
n_{\rm data}=(1-\varepsilon_n)n+o(n).
\]
Thus the recoverable fraction obeys
\[
R_n \ge 1-\varepsilon_n-o(1)\xrightarrow[n\to\infty]{}1.
\]
If Bob accepts, he proceeds to the intended erasure protocol on the data block under the declared work cap. In that case, the evolution on the untested copies is asymptotically the same as in the honest Alice-assisted protocol.

Therefore, under the exclusivity condition and the above test-subset protocol, any dishonest attempt either fails authentication and leaves an asymptotically unit fraction of the memory untouched, or is operationally indistinguishable from the honest strategy on the remaining data block.
\end{proof}

\section{Proof of Theorem 5}

\begin{proof}
In the one-sided semi-device-independent protocol, Bob chooses uniformly at random one of two
dephasing settings $X\in\{R,S\}$ in each run. Over $n$ runs, this produces two blocks:
an $R$-block of size $n_R$ and an $S$-block of size $n_S$, with
\[
n_R+n_S=n,
\qquad
n_R=\frac{n}{2}+o(n),
\qquad
n_S=\frac{n}{2}+o(n)
\]
in the asymptotic limit. The relevant memory in each block is classical, namely Bob’s dephasing
outcome in the chosen basis together with Alice’s classical announcement.

Let
\[
\widetilde W_{A\to B}
=
\frac12\bigl(H(R|A_R)+H(S|A_S)\bigr),
\qquad
\widetilde W_{E\to B}
=
\frac12\bigl(H(R|E_R)+H(S|E_S)\bigr)
\]
denote the honest and adversarial assisted erasure costs, respectively, and assume
\[
\widetilde W_{A\to B}<\widetilde W_{E\to B}.
\]
Set
\[
\Delta := \widetilde W_{E\to B}-\widetilde W_{A\to B} >0.
\]

Choose a sequence $\varepsilon_n\in(0,1)$ such that
\[
\varepsilon_n\to0,
\qquad
n\varepsilon_n\to\infty
\qquad
(n\to\infty).
\]
Bob now chooses at random a test subset of size
\[
m_R=\lfloor \varepsilon_n n_R\rfloor
\quad\text{from the $R$-block,}
\qquad
m_S=\lfloor \varepsilon_n n_S\rfloor
\quad\text{from the $S$-block,}
\]
and leaves the remaining runs untouched.

Since the post-dephasing memory is classical, Bob can apply Shannon compression separately to the
two test blocks, conditioned on Alice’s honest announcements.

For the $R$-test block, the total entropy is asymptotically
\[
m_R\,H(R|A_R),
\]
so classical compression maps it to
\[
m_R\bigl(\log d - H(R|A_R)\bigr)
\]
pure $d$-dimensional registers together with a noisy block carrying the entropy. Likewise, for the
$S$-test block, the number of pure registers is asymptotically
\[
m_S\bigl(\log d - H(S|A_S)\bigr).
\]
Therefore, the total extractable work from the \emph{combined} compressed test blocks in the honest
case is
\begin{align}
W_{\rm ext}^{(A)}
&=
m_R\bigl(\log d - H(R|A_R)\bigr)
+
m_S\bigl(\log d - H(S|A_S)\bigr)
+o(n\varepsilon_n)
\nonumber\\
&=
\varepsilon_n n\bigl(\log d-\widetilde W_{A\to B}\bigr)+o(n\varepsilon_n).
\label{eq:WextA-SDI}
\end{align}

In the dishonest case, Bob applies the same honest compression maps to the $R$- and $S$-test
blocks. These maps need not produce the honest pure/noisy decomposition, but they are still
reversible and therefore preserve the total entropy of each test block.

Hence the total entropy of the compressed test blocks is
\[
m_R\,H(R|E_R)+m_S\,H(S|E_S)
=
\varepsilon_n n\,\widetilde W_{E\to B}+o(n\varepsilon_n).
\]
The total extractable work from the full compressed test block is therefore bounded above by
\begin{equation}
W_{\rm ext}^{(E)}
\le
\varepsilon_n n\bigl(\log d-\widetilde W_{E\to B}\bigr)+o(n\varepsilon_n).
\label{eq:WextE-SDI}
\end{equation}

By \eqref{eq:WextA-SDI} and \eqref{eq:WextE-SDI}, the honest and dishonest extractable works differ by
\[
W_{\rm ext}^{(A)}-W_{\rm ext}^{(E)}
\ge
\varepsilon_n n\Delta + o(n\varepsilon_n).
\]
Since \(\Delta>0\) and \(n\varepsilon_n\to\infty\), for sufficiently large \(n\) the subleading
corrections are smaller than \(\frac{\Delta}{4}\varepsilon_n n\). Bob may therefore choose the
acceptance threshold
\begin{equation}
W_{\rm th}
=
\varepsilon_n n\left(\log d-\frac{\widetilde W_{A\to B}+\widetilde W_{E\to B}}{2}\right),
\label{eq:Wth-SDI}
\end{equation}
so that, for all sufficiently large \(n\),
\[
W_{\rm ext}^{(A)} > W_{\rm th},
\qquad
W_{\rm ext}^{(E)} < W_{\rm th}.
\]

Thus:
\begin{itemize}
    \item in the honest case, the total extracted work from the compressed test blocks exceeds the
    threshold and Bob accepts;
    \item in the dishonest case, no adversarial strategy satisfying the same thermodynamic
    constraints can reach the threshold, so Bob rejects.
\end{itemize}

If Bob rejects, only the tested fraction
\[
m_R+m_S=\varepsilon_n n+o(n)
\]
has been consumed by the authentication step. The remaining
\[
n-(m_R+m_S)=(1-\varepsilon_n)n+o(n)
\]
runs are untouched and therefore remain fully available for future use. Hence the recoverable
fraction obeys
\[
R_n \ge 1-\varepsilon_n-o(1)\xrightarrow[n\to\infty]{}1.
\]

If Bob accepts, he proceeds to the intended erasure protocol on the untested runs under the same
work cap. Therefore, under the exclusivity condition, any dishonest attempt either fails authentication
and leaves an asymptotically unit fraction of the memory untouched, or is operationally
indistinguishable from the honest strategy on the remaining data block. This proves the theorem.
\end{proof}

\end{document}